\documentstyle[12pt,epsf,epsfig,here]{article}
\def\beq{\begin{equation}}
\def\eeq{\end{equation}}
\def\bea{\begin{eqnarray}}
\def\eea{\end{eqnarray}}
\def\bq{\begin{quote}}
\def\eq{\end{quote}}

\def\NP{{\it Nucl.Phys.} }
\def\PL{{\it Phys.Lett.} }
\def\PR{{\it Phys.Rev.} }
\def\PRL{{\it Phys.Rev.Lett.} }

\def\ZP{{\it Z.Phys.} }

\def\gappeq{\mathrel{\rlap {\raise.5ex\hbox{$>$}}
{\lower.5ex\hbox{$\sim$}}}}

\def\lappeq{\mathrel{\rlap{\raise.5ex\hbox{$<$}}
{\lower.5ex\hbox{$\sim$}}}}

\begin{document}
\pagestyle{empty}
\begin{flushright}
CERN-TH/99-136
\end{flushright}
\vspace*{5mm}
\begin{center}
{\bf THE HADRONIC TAU DECAY SIGNATURE OF}\\
{\bf A HEAVY CHARGED HIGGS BOSON AT LHC}
\\
\vspace*{1cm}
{\bf D.P. ROY} \\
\vspace{0.3cm}
Theoretical Physics Division, CERN \\
CH - 1211 Geneva 23 \\
and
\\
Tata Institute of Fundamental Research,\\
Mumbai - 400005, India\\
\vspace*{2cm}
{\bf ABSTRACT}  
\end{center}
\vspace*{5mm}

The hadronic tau decay channel offers by far the best signature for  
heavy charged Higgs boson search at the LHC in the large $\tan\beta$ region. By
exploiting the distinct polarization of the tau and its large transverse mass,
along with the accompanying missing--$p_T$, one can probe for a charged Higgs
boson up to a mass of about 600 GeV in an essentially background-free
environment. The transverse mass distribution of the tau jet also provides a
fairly unambiguous estimate of the charged Higgs boson mass.

\vspace*{1cm}

\vspace*{2.5cm}
\begin{flushleft}
CERN-TH/99-136 \\
May 1999
\end{flushleft}

\vfill\eject
%\pagestyle{empty}
%\clearpage\mbox{}\clearpage

\setcounter{page}{1}
\pagestyle{plain}

The charged Higgs boson carries the hallmark of a two Higgs doublet model and
in particular the minimal supersymmetric Standard Model (MSSM). The two complex
doublets correspond to eight scalar states, three of which are absorbed as
Goldstone bosons. This leaves five physical particles -- the two neutral
scalars, a pseudoscalar and a pair of charged Higgs bosons $H^\pm$ \cite{aaa}.
While one of these neutral Higgs bosons may be hard to distinguish from that of
the Standard Model, the $H^\pm$ carries a distinctive signature of the
supersymmetric Higgs sector. Therefore the charged Higgs boson search is of
considerable importance in probing the SUSY Higgs sector.

The charged Higgs boson couplings to fermions are given, in the diagonal 
Cabibbo-Kobayashi-Maskawa matrix approximation, by
\beq
{\cal L} = {g\over \sqrt{2} m_W}~~H^+ \left\{ \cot \beta~ m_{u_i} \bar u_i
d_{i_L} + \tan\beta~ m_{d_i} \bar u_i d_{iR} + \tan\beta~ m_{\ell_i} \bar
\nu_i
\ell_{iR}\right\} + H.C.~,
\label{one}
\eeq
where $i$ denotes the generation  index and $\tan\beta$ represents the ratio of
the vacuum expectation values of the two Higgs doublets. The QCD corrections
are taken into account, in the leading log approximation, by substituting the
quark mass parameters by their running masses evaluated at the $H^\pm$ mass
scale \cite{bb}.

It is clear from (\ref{one}) that, for a charged Higgs boson lighter than the
top quark, 
\beq
H^+\rightarrow \tau^+\nu
\label{two}
\eeq
is the dominant decay mode over the large $\tan\beta$ region. This is in
striking constrast to the universal decay branching fractions of $W$ into each
fermion pair. This difference has been utilized to search for $H^\pm$ in the
top quark decay data of the CDF experiment in the $\tau$ channel at the Tevatron
and obtain modest limits on the $H^\pm$ mass in the large $\tan\beta$ region
\cite{cc}. 
The search can be extended up to $m_H \simeq m_t$ and over a wider region of
$\tan\beta$ at the upgraded Tevatron  and LHC, particularly by exploiting
the opposite states of $\tau$ polarization coming from charged Higgs and $W$
boson decays \cite{dd},\cite{ee}.

But the search for a heavy charged Higgs boson, $m_H > m_t$, at a hadron
collider is generally considered very hard, because in this case the dominant
decay mode
\beq
H^+ \rightarrow t \bar b~,
\label{three}
\eeq
suffers from a large QCD background. The dominant production process for a
heavy $H^\pm$ at the LHC is its associated production with a top quark via
gluon-$b$ quark fusion
\beq
g\bar b \rightarrow H^+\bar t
\label{four}
\eeq
or the related process of gluon-gluon fusion
\beq
gg \rightarrow H^+ \bar t b~.
\label{five}
\eeq
Integrating out the kinematic variables of the final state $b$ quark in the
second case results in a similar size of cross-section (within a factor of 2)
as the first. This is as expected since $m_b \ll m_t, m_H$. The prospect of a
heavy $H^\pm$ search at the LHC was considered in \cite{ff},\cite{ggg} by
combining the associated production of $H^\pm$ and $t$, with its dominant decay
mode (\ref{three}). The major background comes from the QCD processes of $t\bar
t b$ or $t\bar t q (g)$, where the light quark or gluon jet could be
misidentified as $b$. One could find a viable signature for the $H^\pm$ via
triple $b$ tagging and reconstruction of $t, \bar t$ and $H^\pm$ masses; but
only over a limited range of $\tan\beta$, i.e.,
\beq
\tan\beta \sim 1 \quad\quad {\rm or} \quad\quad \tan\beta \sim {m_t\over m_b}~.
\label{six}
\eeq
The reason why these two regions of $\tan\beta$ are favoured is the enhancement
of the $t\bar b H^\pm$ coupling and the resulting signal cross-section as one
can clearly see from (\ref{one})\footnote{Interestingly these two regions of
$\tan\beta$ are favoured by GUT scale unification of $b$ and $\tau$ masses for
a related reason; i.e., one needs a large Higgs Yukawa coupling in the RGE to
control the growth of the $b$ mass at the low energy scale \cite{hh}.}. More
recently the analysis has been extended to the 4 $b$-tagged channel \cite{jj},
by combining the associated production of $H^+\bar t b$ (\ref{five}) with the
dominant decay mode (\ref{three}). Unfortunately, it requires the associated
$b$ quark in (\ref{five}) to be hard, which reduces the signal cross-section
by a large factor. Consequently this signature is again viable only over the
limited $\tan\beta$ regions of Eq. (\ref{six}). It may be added here that the
associated production of $H^\pm$ and $W^\mp$ bosons has been considered
recently in \cite{kk},\cite{lll}. Being a second-order electroweak process,
however, the size of the resulting signal is much smaller than (\ref{four}),
while it suffers from the same background. Consequently, it does not seem to
offer a viable signal at the LHC \cite{lll}.

The purpose of this work is to try to construct a viable signature for heavy
$H^\pm$ search at the LHC via its $\tau$ decay mode (\ref{two}). The importance
of this decay mode for heavy $H^\pm$ search at the LHC has been recently
emphasized in \cite{mm}. However, there is as yet no quantitative analysis of
this signal  along with the relevant background. For the signal we shall
consider the associated production of $H^\pm$ 
with $t$ via $gb$ fusion followed by its $\tau$ decay, i.e.,
\beq
g\bar b \rightarrow H^+ \bar t~;~~~H^+\rightarrow \tau^+\nu~,~~~ \bar t
\rightarrow \bar b q\bar q
\label{seven}
\eeq
and its charge conjugate process. The major background comes from the $t\bar t$
production, i.e.,
\beq
gg(q\bar q) \rightarrow t\bar t~,~~~ t \buildrel{W}\over{\rightarrow} b
\tau^+\nu~,~~~ \bar t \rightarrow \bar b q\bar q
\label{eight}
\eeq
and the charge conjugate final state. Even with a 50\% $b$-tagging efficiency,
the second $b$ in the background will escape identification most of the time;
and this will be impossible to distinguish from a QCD jet accompanying the
signal in the LHC environment. Thus we have to start with a signal/background
ratio, which is even smaller than the $t\bar t b $ channel, considered in
\cite{ff}, \cite{ggg}.

As we shall see below, however, one can kinematically separate the signal from
the background in this case by exploiting the hardness of the $\tau$-jet, its
distinctive polarization and finally its large transverse mass with the
accompanying missing-$p_T~~(p\llap{$/$}_T)$. It is to preserve this last feature
that we have considered the hadronic decay of the accompanying top in
(\ref{seven}) and (\ref{eight}).

The basic signal cross-section is obtained by multiplying the gluon-$b$ quark
fusion cross-section with the hadronic BR of top decay (= 2/3) along with the
BR of $H\rightarrow\tau\nu$, i.e.,
\beq
B_\tau = \left[ 1 + {3m^2_b\over m^2_\tau}~~\left( 1 + {m^2_t\over
m^2_b\tan^4\beta}\right)~~\left( 1 - {m^2_t\over m^2_H}\right)^2\right]^{-1}~.
\label{nine}
\eeq
As mentioned earlier, here $m_b$ stands for the running mass at the $H^\pm$
mass scale. It is adequate for our purpose to take \cite{nn}
\beq
m_b (m_H) \simeq m_b (m_t) \simeq 2.8~{\rm GeV}
~.
\label{ten}
\eeq
The $m_t$ in the first factor of (\ref{nine}) comes from Yukawa coupling and in
the second factor from phase space. The corresponding values should correspond
to the running and the pole masses respectively, which differ by 5\%. For
simplicity we shall put $m_t$ = 175 GeV for both and $m_\tau$ = 1.78 GeV. The
resulting branching fraction is $B_\tau$ = 0.7 at $m_H$ = 200 GeV, going down
to 0.15 at $m_H$ = 500 GeV, over the large $\tan\beta~ ( \gappeq$ 10) region of
our interest.

We shall consider the hadronic decay channel of $\tau$, which carries a better
imprint of its hardness and state of polarization, while the leptonic
$(e,\mu)$ decay of $\tau$ from a 200-300 GeV $H^\pm$ will look similar to the
direct leptonic decay of a $W$ boson. We shall concentrate on the 1-prong
hadronic decay of $\tau$, which  is best suited to $\tau$ identification. It
accounts for 80\%   of hadronic $\tau$ decay and 50\% of its total decay
width. The main contributors to 1-prong hadronic $\tau$ decay are
\bea
\tau^\pm &\rightarrow& \pi^\pm \nu_\tau~~ (12.5 \%)~, \\
&& \nonumber \\
%\label{eleven}
\tau^\pm &\rightarrow& \rho^\pm \nu_\tau \rightarrow \pi^\pm \pi^0 \nu_\tau
~~(26
\%)~,\\
&& \nonumber \\
%\label{twelve}
\tau^\pm &\rightarrow& a_1^\pm \nu_\tau \rightarrow \pi^\pm\pi^0 \nu_\tau~~ (7.5
\%)~,
\eea
where the branching fractions for $\pi$ and $\rho$ channels include the small
$K$ and $K^*$ contributions respectively \cite{oo}, which have identical
polarization effects. Together they account for a little over 90\% of the
1-prong hadronic decay of $\tau$. The CM angular distribution of $\tau$ decay into
$\pi$ or a vector meson $v ~(=\rho , a_1)$ is simply given in terms of its
polarization as \cite{dd}, \cite{ee}
\bea
{1\over\Gamma_\pi} ~\cdot~ {d\Gamma_\pi\over d\cos\theta} &=& {1\over 2}~~(1 +
P_\tau \cos\theta)~, \\
&&  \nonumber \\
{1\over\Gamma_v} ~\cdot~ {d\Gamma_{vL}\over d\cos\theta} &=& {{1\over
2}m^2_\tau\over m^2_\tau + 2m^2_v}~~(1 + P_\tau \cos\theta)~, \\
&&  \nonumber \\
{1\over\Gamma_v} ~\cdot~ {d\Gamma_{vT}\over d\cos\theta} &=& {m^2_v\over
m^2_\tau + 2m^2_v}~~(1 - P_\tau \cos\theta)~, 
\eea
where $P_\tau$ = +1 (-1) for the $\tau$ coming  from $H^\pm~(W^\pm)$ decay and
$L, T$ denote the longitudinal and transverse polarization states of the vector
meson. This angle is related to the fraction $x$ of $\tau$ lab. momentum
carried by its decay meson, i.e., the (visible) momentum of the $\tau$-jet, via
\beq
\cos\theta = {2x -1 -m^2_{\pi,v}/m^2\tau \over
1 - m^2_{\pi,v}/m^2_\tau}~.
\label{seventeen}
\eeq
It is clear from Eqs. (14)-(17) that the effect of $\tau$ polarization is to
give a harder $\tau$-jet from the signal relative to the background for the
$\pi$ and the longitudinal vector meson contributions; but it is the other way
round for the transverse part. The latter results in diluting the polarization
effect by 50\%  for $\rho$ and practically washing it out for the $a_1$
contribution.

It is possible to suppress the transverse $\rho$ and $a_1$ contributions and
enhance the difference between the signal and the background in the 1-prong
hadronic $\tau$ decay channel even without identifying the individual mesonic
contributions to this channel. The key feature of vector meson decays, relevant
to this purpose, is that the transverse $\rho$ and $a_1$ favour even sharing
of momentum among their decay pions, while the longitudinal $\rho$ and $a_1$
decays favour uneven distributions, where the charged pion carries either very
little or most of the momentum. It is easy to derive this quantitatively for the
$\rho$ decay. But one has to assume a dynamical model for the $a_1$ decay to get
a  quantitative result. We shall assume the model of Ref. \cite{pp}, based on
conserved axial vector current approximation, which provides a good descriptin
of the $a_1 \rightarrow 3\pi$ data. A detailed
account of the $\rho$ and $a_1$ decay formalisms can be found in \cite{dd},
\cite{ee}, along with the distributions of the resulting ratio,
\beq
r = {p_{\pi^\pm} \over p_{\tau -jet}}~,
\label{eighteen}
\eeq
for the longitudinal and transverse $\rho$ and $a_1$. It suffices to state the
main result here, i.e., this ratio is peaked near 0 and 1 for the
longitudinal parts, but at the middle for the transverse parts. This suggests two
ways of suppressing the transverse $\rho$ and $a_1$ contributions to the
1-prong hadronic $\tau$ decay channels \cite{ee}. The first one is to demand
\beq
r = {p_{\pi^\pm}\over p_{\tau - jet}} > 0.8~,
\label{nineteen}
\eeq
i.e., over 80\% of the $\tau$-jet $p_T$ is carried by the charged track. It
retains only the $\pi$ and half of $\rho_L$ contribution, while sacrificing the
other half of $\rho_L$  and $a_{1L}$ along with the $\rho_T$ and $a_{1T}$. But
it gives a robust signature of a very hard single track, accounting for over
80\% of the calorimetric energy deposit. The second method is to demand a hard
distribution in 
\beq
\Delta p_T = \vert p_T^{\pi^\pm} - p_T^{\pi^0}\vert~,
\label{twenty}
\eeq
i.e., the difference of momenta carried by the charged track and the
accompanying neutral pion(s) instead of their sum. It retains most of the
$\rho_L$ and $a_{1L}$ contributions along with the $\pi$, since it includes
both $r \simeq 1$ and $r \simeq 0$ regions. However, the latter corresponds to
a soft charged track along the direction of a hard calorimetric energy deposit,
which could be deflected away from this direction by the magnetic field. Note
that this part is present even in the normal $p_T$ distribution of the 1-prong
$\tau$-jet, but not the one satisfying (\ref{nineteen}). Hopefully this problem
will be taken care of in a more sophisticated analysis including detector
simulation. For the present purpose we shall illustrate the enhancement of
signal/background ratio via both  methods, without worrying about this
problem.

Our results are based on a parton level Monte Carlo calculation of the signal
and background processes (\ref{seven}) and (\ref{eight}), followed by the
1-prong hadronic decay of $\tau$ via $\pi, \rho$ and $a_1$. The LO
cross-sections for the signal and background processes are convoluted with the
LO parton distributions of CTEQ 4L \cite{qq} using the threshold energies,
$(m_H + m_t)$ and $2 m_t$, as the scale parameter. We have checked that the
resulting cross-sections are 10-20\% higher than those obtained with the NLL
parton distributions of MRS (G) \cite{rr}. The parton distributions were used
via the CERN pdflib version 7.09.

We have tried to simulate detector resolution by a gaussian smearing of $p_T$,
with $(
\sigma (p_T)/p_T)^2 = (0.6/\sqrt{p_T})^2 + (0.04)^2$, for all the jets including
the $\tau$-jet. The $p\llap{$/$}_T$ is obtained by vector addition of all these
$p_T$'s after the gaussian smearing. As a basic set of selection cuts
we require
\beq
p_T > 20~{\rm GeV~~~and}~\eta < 3
\label{twentyone}
\eeq
for all the jets, where $\eta$ denotes pseudorapidity. Only the $b$-jet
accompanying $\tau$ in the background process (\ref{eight}) is not subject to
this constraint. We also require a minimum separation of
\beq
\Delta R = [(\Delta\phi)^2 + (\Delta\eta)^2]^{1/2} > 0.4
\label{twentytwo}
\eeq
between each pair of jets, where $\phi$  denotes the azimuthal angle. The jets
are required to satisfy the $W$ and top mass constraints, i.e., 
\beq
\vert m_{jj} - m_W\vert < 15~{\rm GeV}~,~~~\vert m_{jjj} - m_t\vert < 25 ~{\rm
GeV}~.
\label{twentythree}
\eeq
With the above resolution smearing, these mass constraints are satisfied
essentially without any loss to the signal or background.
We shall also require the third jet, entering the top mass constraint, to be
$b$-tagged for an unambiguous identification of the accompanying top. The
signal and background cross-sections presented below are to be multiplied by
the corresponding efficiency factor, which may be optimistically assumed to be
$\sim$ 0.5.

\begin{table}
\caption{Charged Higgs signal and background cross-sections in the 1-prong
hadronic $\tau$ plus multijet channel for $\tan\beta = 40$. Here $p_T$ refers
to the visible $p_T$ of the $\tau$, while $p_T^c$ and $\Delta p_T$ are defined
in the text. The last row shows the $\tau$ branching fractions of the signal
and the background, which are included in these cross-sections.}
\label{Table1}
\begin{center}
\begin{tabular}{|l|c|c|c|c|}\hline
$\sigma$ (fb) & $H^+$ (200 GeV) & $H^\pm$ (400 GeV) & $H^\pm$ (600 GeV) & $Bg$
\\ \hline
Basic Cuts & 580 & 30.1 & 6.7 & 16176 \\
(Eqs. 21-23) &&&& \\ \hline
$p_T >$ 100 GeV & 132 & 17.7 & 5.1 & 720 \\ \hline
$p^c_T >$ 100 GeV & ~~~61.1 & ~~8.1 & 2.3 & 124 \\ \hline
$\Delta p_T >$ 100 GeV & ~~~68.6 & 10.7 & 3.6 & 136 \\ \hline
$B_\tau$ & ~~~~~~0.71 &         ~~~0.17 & ~~0.14 & 2 $\times$ 0.11 
\\ \hline
\end{tabular}
\end{center}
\end{table}

Table 1 shows the signal cross-sections for $m_H$ = 200, 400 and 600 GeV at
$\tan\beta = 40$ along with the $\bar tt$ background. The cross-sections,
obtained with the basic selection cuts of Eqs.~(21)-(23), are shown in the
first row. At this stage the background is 2 to 3 orders of magnitude larger
than the signal. The second row shows the cross-sections after a 
\beq
p_T > 100~{\rm GeV}
\label{twentyfour}
\eeq
cut on the $\tau$-jet. As expected, it reduces the background more severely than
the signal, so that it is now 1 to 2 orders of magnitude larger than the
signal. The signal/background ratio can be improved further by exploiting the
$\tau$ polarization effect as discussed above. The third row shows the
cross-section for 
\beq
p_T^c > 100~{\rm GeV}~,
\label{twentyfive}
\eeq
where $p^c_T$ denotes the $p_T$ of the $\tau$-jet satisfying (19), i.e., where
the charged track carries over 80\% of the $p_T$. It reduces the background by
a factor of 6, while costing a factor of 2 to the signal. The fourth row shows
the cross-sections for 
\beq
\Delta p_T > 100~{\rm GeV}~.
\label{twentysix}
\eeq
The signal and background cross-sections in this case are similar to the
previous case.

Figures 1a,b,c show the distributions of the signal and background
cross-sections against $p_T, p^c_T$ and $\Delta p_T$ respectively. As expected,
the signal cross-sections get harder with increasing $H^\pm$ mass. One can also
see the hardness of the signal cross-section relative to the background
increasing as one goes from $p_T$ to $p^c_T$ or $\Delta p_T$. Thus the $p_T$
distribution of the background is seen to remain higher than the signal over
the entire range; but the $p^c_T$ distribution of the 200~GeV $H^\pm$ signal
overtakes the background, and even the 400 and 600 GeV $H^\pm$ signals catch
up with the latter. The same is true for the $\Delta p_T$ distributions.
However, the distribution in $p^c_T$ may be of greater significance in view of
the robustness of this variable.

Figure 2 shows the signal and background cross-sections, with $p^c_T >$ 100 GeV
(25), against the azimuthal angle of the $p\llap{$/$}_T$ relative to the
$\tau$-jet. The background $(W\rightarrow\tau\nu)$ events, satisfying the large
$p^c_T$ cut, necessarily require a large $p_T$ boost for the $W$ boson. This
results in a small opening angle between its decay products. In contrast the
$H^\pm\rightarrow\tau\nu$ events require only a modest $p_T$ boost for $M_H$ =
200 GeV and none for the larger $H^\pm$ masses. This accounts for the flat
azimuthal opening angle distribution between the $\tau$ and 
$p\llap{$/$}_T$ for the 200 GeV $H^\pm$ signal and backward peaks for the
higher masses. One gets similar results for the signal and background
cross-sections with $p_T$ or $\Delta p_T$ cuts of 100 GeV.

Similarly the 
$p\llap{$/$}_T$ distribution of the signal is harder than the background; and
the difference increases with increasing $H^\pm$ mass. The quantity in which
the cumulative difference between the signal and the background becomes most
striking is the transverse mass of the $\tau$-jet with the $p\llap{$/$}_T$,
i.e.,
\beq
m_T = 2p_T p\llap{$/$}_T (1-\cos\phi),~~m^c_T = 2 p^c_T p\llap{$/$}_T
(1-\cos\phi),
\label{twentyseven}
\eeq
since it is kinematically constrained to be smaller than the parent $H^\pm~(W)$
mass. The $m_T$ and $m^c_T$ distributions of the signal and background
cross-sections are shown in Fig. 3a,b, which include the $p_T~(p^c_T) >$ 100 GeV
cut. There is a clear separation between the signal and the background
cross-sections, particularly in Fig. 3b. It thus  seems possible to
practically eliminate the background via a suitable $m^c_T$ cut with very
little cost to the signal cross-section. Consequently one can probe for the
$H^\pm$ signal in an essentially background-free environment. It is also
evident from Fig. 3a,b that the transverse mass distribution can provide a
fairly unambiguous estimate of the $H^\pm$ mass. In comparing Fig. 3a with 3b,
one sees a significantly larger background in the former case, resulting in a
larger encroachment into the signal region. Thus the separation of the signal
and background in the
$m_t$ distribution may be less clean, particularly if the uncertainty in the
$p\llap{$/$}_T$ turns out to be larger than that suggested by the assumed
resolution smearing. On the other hand, if it is possible to separate the two
from the $m_T$ distribution, then the polarization effect can be used as a
confirmatory test of the signal. We have checked that the background
cross-section at large $m_T ~(>$ 100 GeV) is dominated by the transverse $\rho$
and
$a_1$ contributions. Consequently its distribution in the ratio $r$ (18)
vanishes at $r \simeq 0$ and 1 while peaking in the middle. In contrast the
signal is dominated by the $\pi, \rho_L$ and $a_{1L}$ contributions, showing
peaks at
$r \simeq 0$ and 1, and a dip in the middle.

Assuming that the signal can be clearly separated from the background from the
$m_T$ or $m^c_T$ distribution, the $H^\pm$ discovery limit of the LHC will be
primarily determined by the signal size. The size of the signal cross-sections
shown in Fig. 3a,b corresponds to those listed in the second and third rows of
Table 1. They have to be multiplied by the $b$-tagging efficiency
of $\sim 0.5$. Still one expects a signal size of at least 1 fb up to $m_H$ =
600 GeV. This will correspond to at least 100 events with the expected annual
luminosity of 100 fb$^{-1}$ for the high luminosity run of the LHC -- i.e., a
total of at least 300-400 events over a period of 3-4 years. Of course we have
concentrated so far on a favourable value of $\tan\beta$ = 40. 
For lower values of $\tan\beta$ the signal cross-section is expected to go down
like $\tan^2\beta$. Thus one expects a total of at least 100 (25) signal events
at $\tan\beta$ = 20 (10), for a $H^\pm$ mass going up to 600 GeV. Hence the
hadronic $\tau$ decay channel seems to offer a viable signature for the $H^\pm$
search at the LHC up to 600 GeV over a sizeable part of the large $\tan\beta$
region, i.e., $\tan\beta \gappeq 10$.

In summary, we have analysed the prospect of a heavy $H^\pm$ search at the LHC
via its hadronic $\tau$ decay channel. The dominant signal comes from the
associated production of $H^\pm$ with a top quark, while the dominant background
is from top pair production. Although the signal is 2 to 3 orders of magnitude
smaller than the background, it can be effectively separated by exploiting the
hardness and the distinctive polarization of the $\tau$ coming from $H^\pm$
decay. In particular an effective method of utilizing the polarization effect
is to select the 1-prong hadronic $\tau$ jets, where the charged prong carries
over 80\% of its visible momentum. The resulting transverse mass of the $\tau$
jet with the accompanying $p\llap{$/$}_T$ shows a clear separation between the
signal and the background, as well as providing a fairly unambiguous estimate of
the $H^\pm$ mass. Thus the viability of the signal is primarily determined by
the signal size. For the high luminosity run of the LHC, one expects a viable
signal up to the $H^\pm$ mass of 600 GeV over the large
$\tan\beta ~(\gappeq$~10) region.

I thank Daniel Denegri for persuading me to undertake this investigation. I
also thank Michelangelo Mangano, Debajyoti Choudhury and Sunanda Banerjee for
discussions and Swagata Banerjee for help with the figure-plotting programme.
This work was partly supported by the IFCPAR under project No. 1701-1 on
Collider Physics.

\vfill\eject

\vfill\eject

\noindent

%--------------------------------------------------------------------%
\begin{figure}[htb]
% \vspace*{-1.8cm}
\vspace*{2ex}
\hspace*{-0.8cm}
\centerline{
\epsfxsize=6.cm\epsfysize=7.0cm
\epsfbox{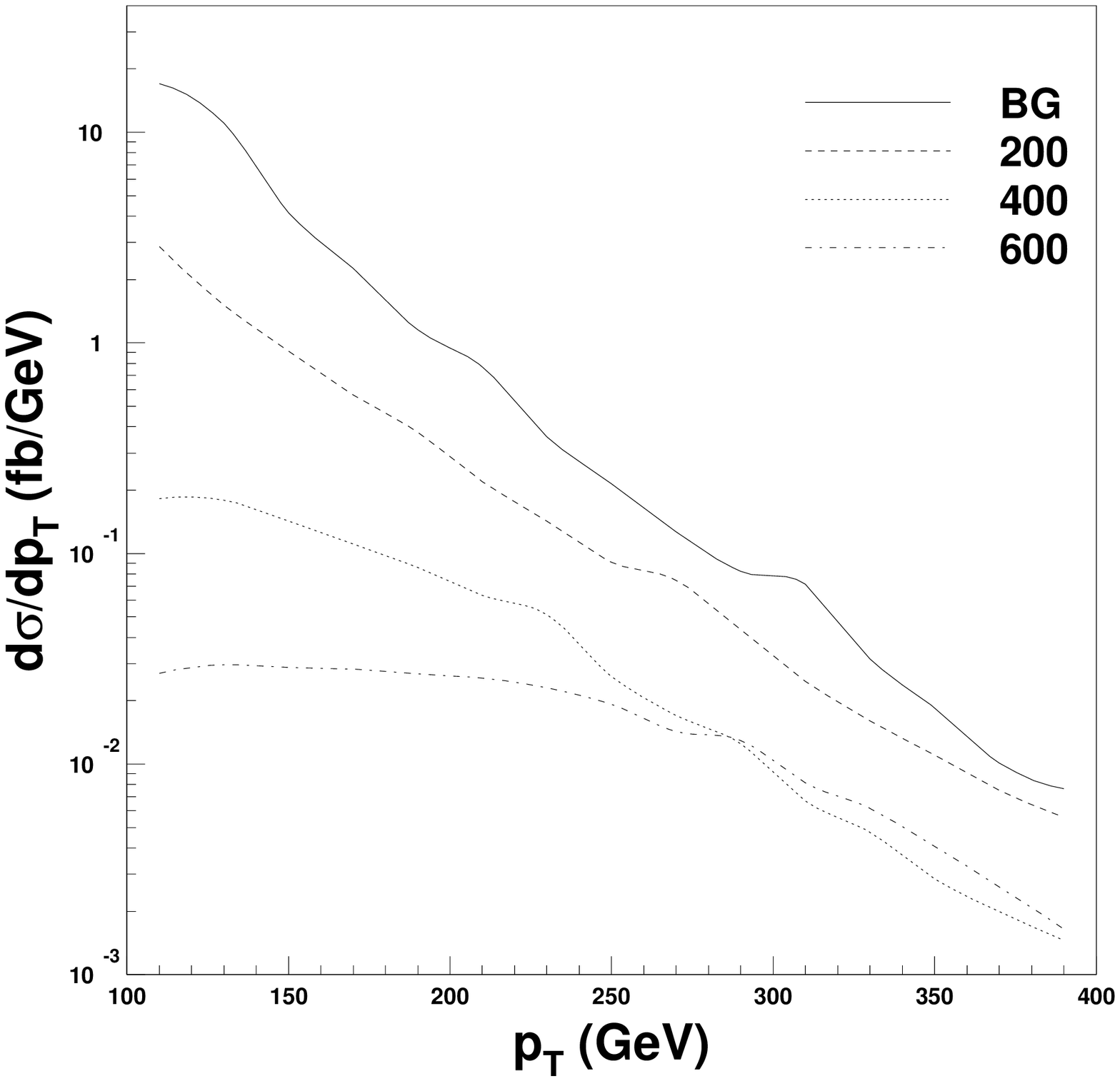}
\vspace*{-0.0cm}
\hspace*{-0.5cm}
\epsfxsize=6.cm\epsfysize=7.0cm
\epsfbox{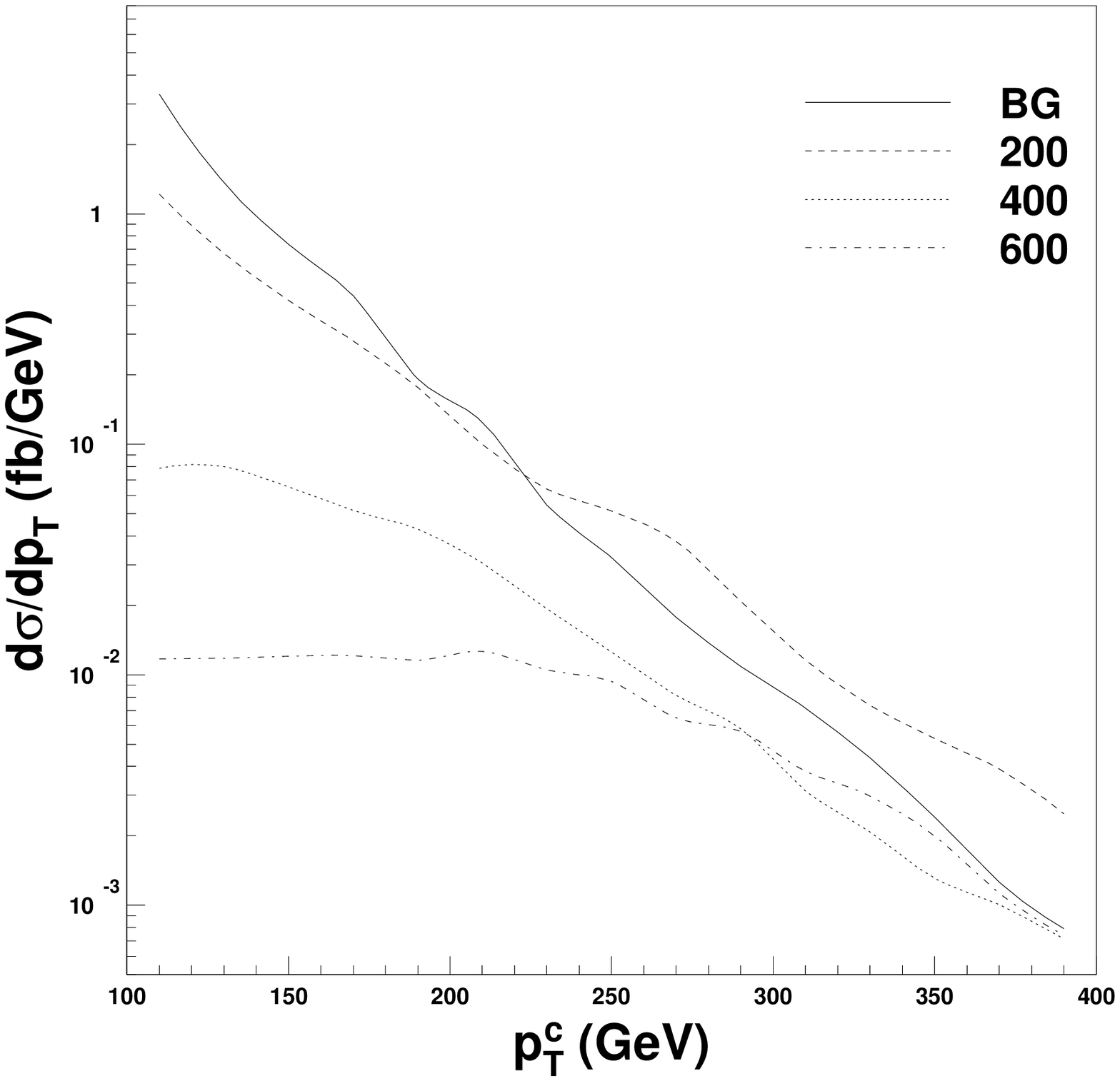}
\vspace*{-0.0cm}
\hspace*{-0.5cm}
\epsfxsize=6.cm\epsfysize=7.0cm
\epsfbox{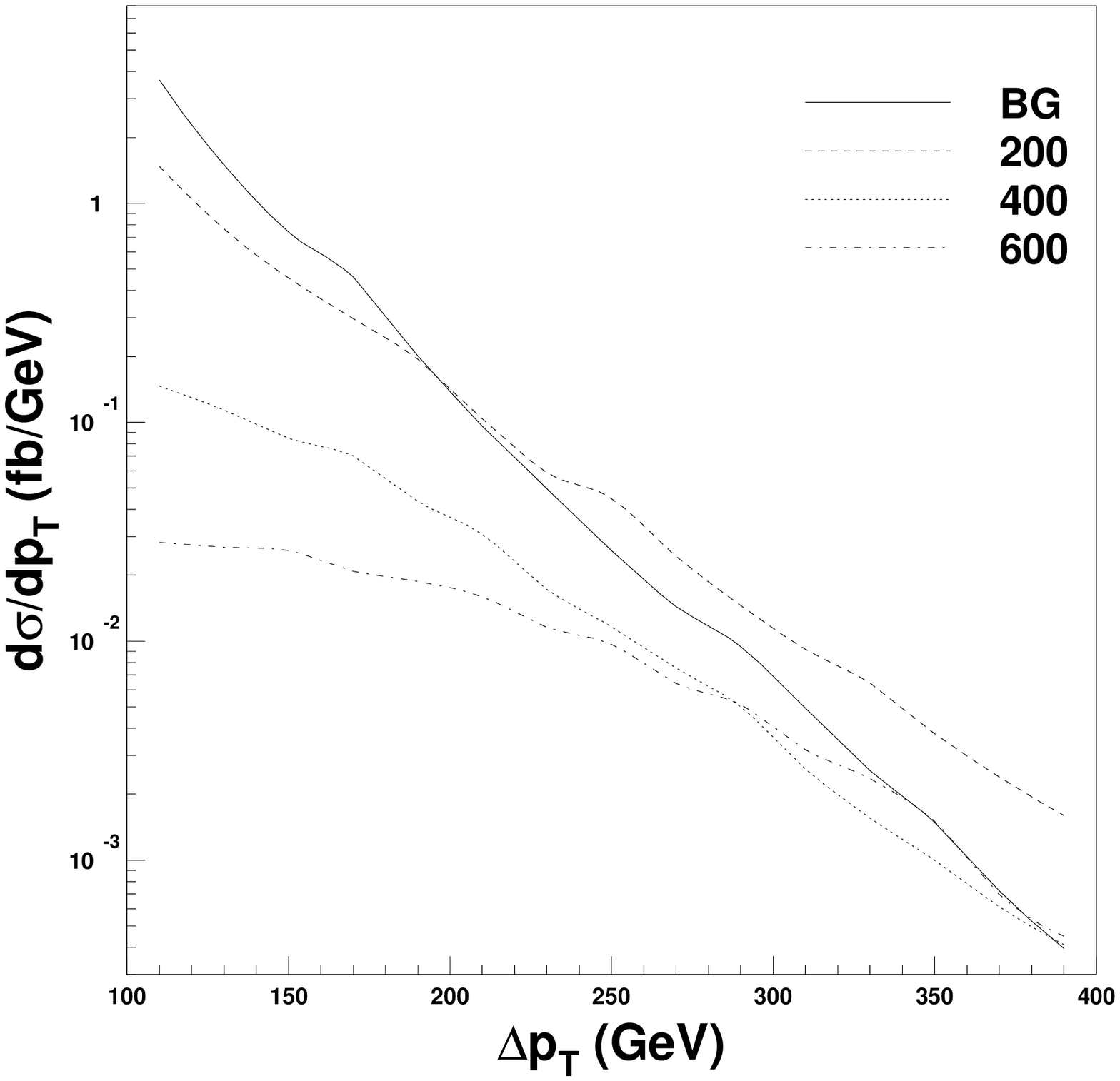}
\vspace*{-0.3cm}
}
\caption[fig:fig1]{Distributions of the $H^\pm$ signal and the background
cross-sections in a) $p_T$ of the $\tau$-jet, b) $p_T$ of those $\tau$-jets
where the charged pion carries over 80\% of the $p_T$, c) difference of the
charged and neutral pion contributions to the $p_T$ of the $\tau$-jet. The
signal cross-sections in this and the following figures correspond to
$\tan\beta = 40$.
}   %%%%%%% \label{fig:3}
\end{figure}

%--------------------------------------------------------------------%
\begin{figure}[htb]
% \vspace*{-1.8cm}
\vspace*{2ex}
\hspace*{-0.8cm}
\centerline{
\epsfxsize=7.cm\epsfysize=7.0cm
\epsfbox{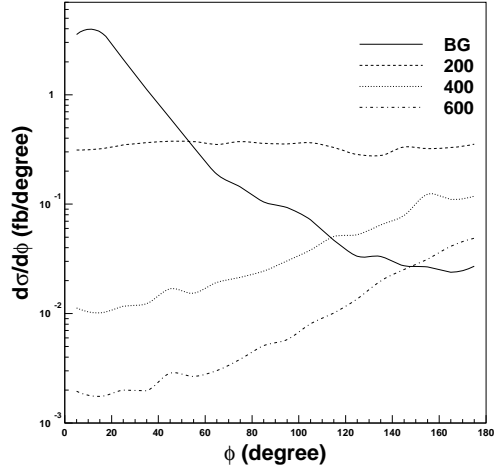}
}
\caption[fig:fig2]{Distribution of the $H^\pm$ signal and background
cross-sections, satisfying $p^c_T >$ 100~GeV, in the azimuthal angle between
the $\tau$-jet and the missing-$p_T$.
}   %%%%%%% \label{fig:3}
\end{figure}
%--------------------------------------------------------------------%
\begin{figure}[htb]
% \vspace*{-1.8cm}
\vspace*{2ex}
\hspace*{-0.8cm}
\centerline{
\epsfxsize=7.cm\epsfysize=7.0cm
\epsfbox{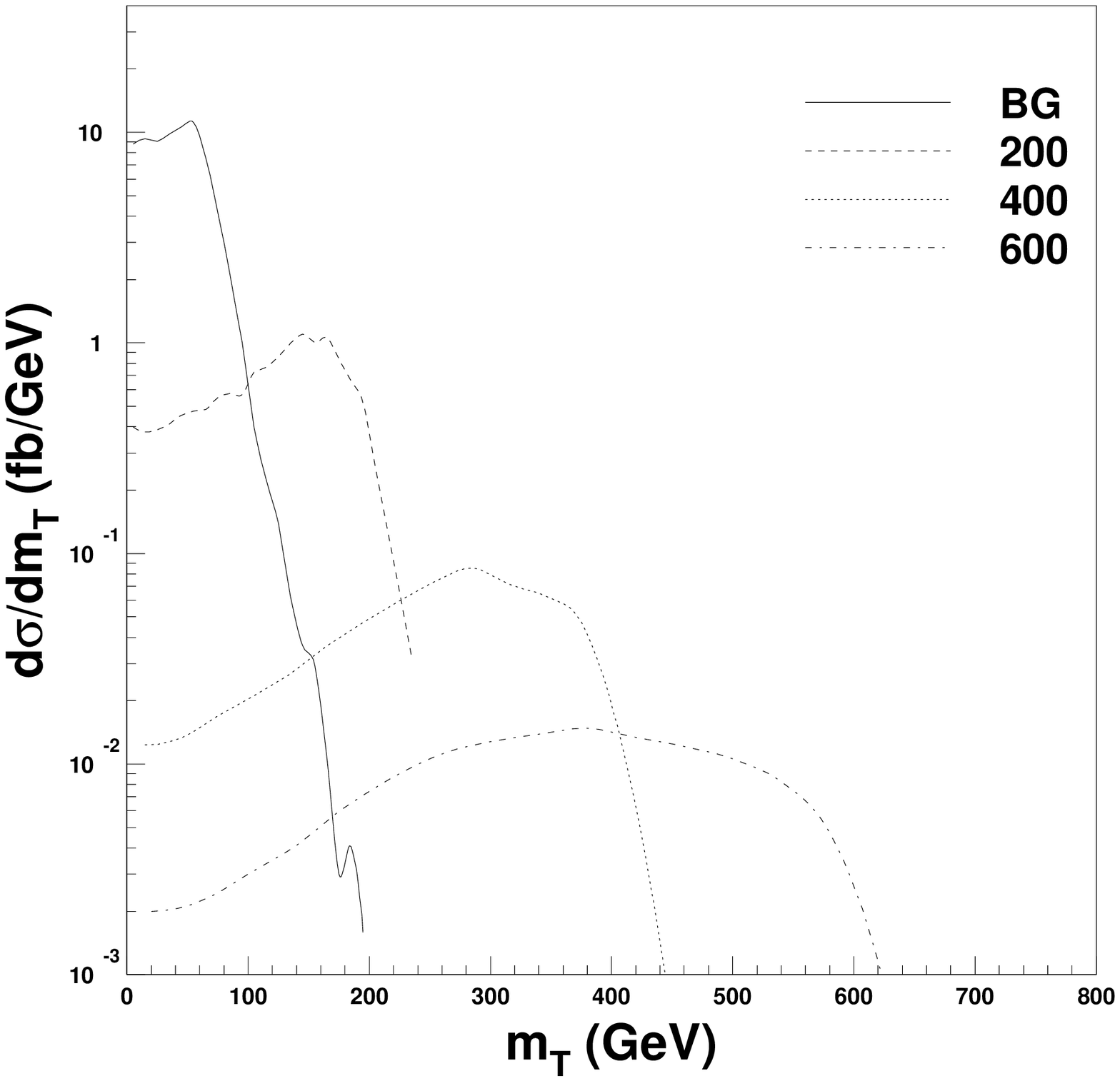}
\vspace*{-0.0cm}
\hspace*{-0.5cm}
\epsfxsize=7.cm\epsfysize=7.0cm
\epsfbox{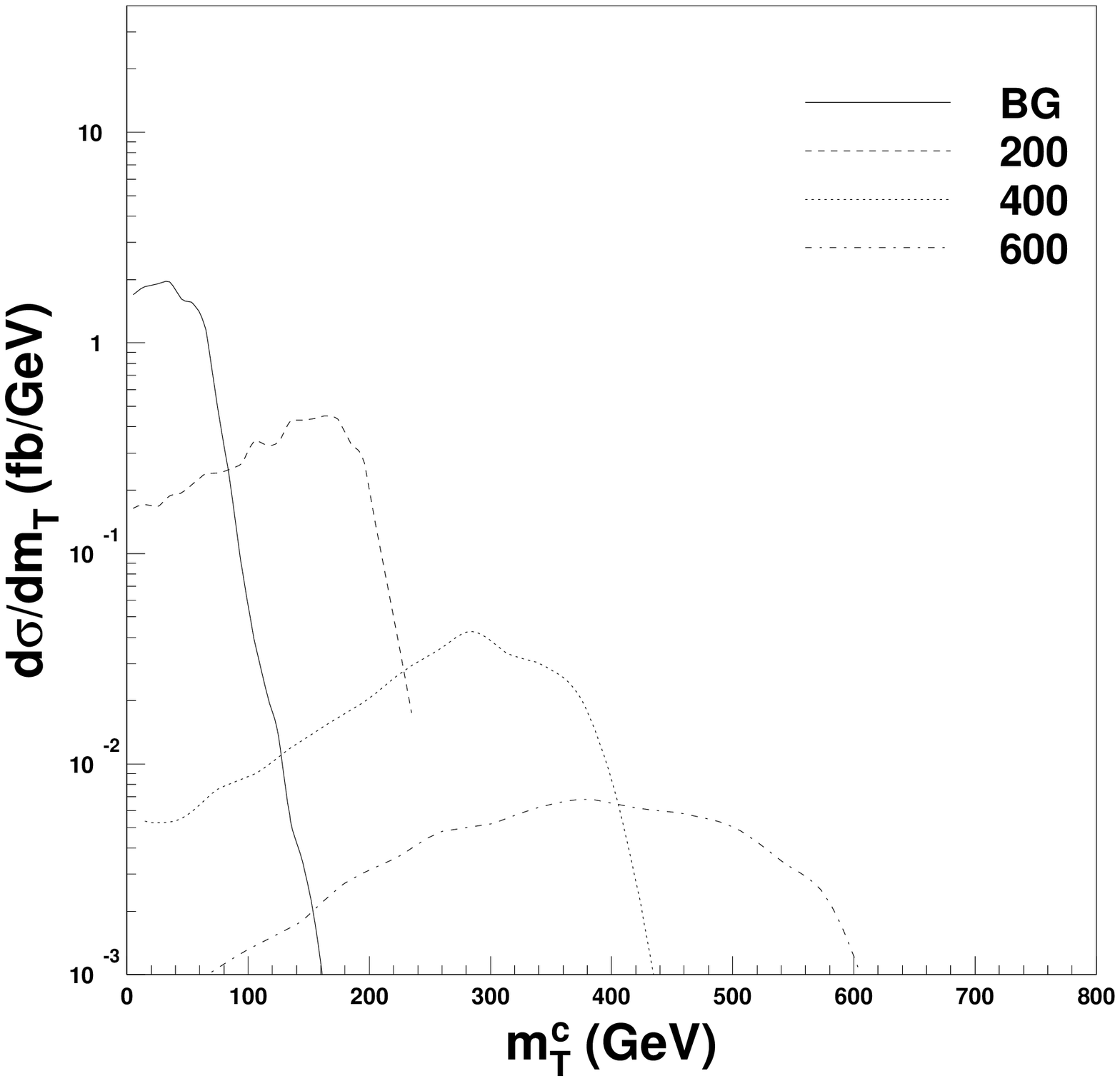}
\vspace*{-0.3cm}
}
\caption[fig:fig3]{Distribution of the $H^\pm$ signal and background
cross-sections in the transverse mass of the $\tau$-jet with the missing-$p_T$
for a) all 1-prong $\tau$-jets, b) those where the charged pion carries over
80\% of the $\tau$-jet $p_T$.
}   %%%%%%% \label{fig:3}
\end{figure}

\end{document}